# A Framework for Assessing Universal Service Obligations: A Developing Country Perspective[1]

Rekha Jain
Pinaki Das

Rekha @iimahd.ernet.in
Indian Institute of Management, Ahmedabad
India.

## Introduction

A critical element of most national telecom policy objectives is advancing universal service. International Telecommunications Union (ITU) considers universal service provision as purely for the purpose of making available telecommunication services of specified quality, and in light of the specific national conditions, at an affordable price to that minority of potential users who do not have the service and for whom not having access to the service would constitute a social or economic disadvantage[2].

Besides the political compulsions for providing universal service, there are economic arguments relating to network externalities and economic development. Increasing the network reach has positive effect. Therefore, governments normally place a Universal Service Obligation (USO), the obligation on one or more operators to provide universal service. Uneconomic USO means the provision of such services to those users whom an operator would not serve if it were to apply its normal commercial criteria of profitability. The underlying aim of universal service is to ensure that the benefits of cheaper and better quality telephone service and other benefits of increased competition and choice are passed on to all users. Key issues that regulators and policy makers face with regard to universal service provision are estimation and mechanisms for funding, mechanism for deployment and involvement of private sector, and monitoring targets (Crandall and Waverman, 2000).

This paper highlights the various issues in USO from a developing country perspective. The first part of this paper gives an overview of USO practices and issues. The second part reviews the Telecom Regulatory Authority of India's recommendations on USO cost estimation. In the third part, the paper analyzes characteristics of rural exchanges with a view to evolve a framework for assessing USO. This framework is applicable to developing countries as the study carried out in this paper is in the context of low telecom penetration and non-availability of data with regulators.

---

[1] The authors acknowledge the help and co-operation extended to us by the Chief General Manager, Mr. PK Chanda, Gujarat Telecom Circle and his staff.

[2] http://www.itu.int

# Issues in USO

## Definition of USO: Differences in Developed and Developing Countries

What constitutes basic services varies from country to country depending on the economic conditions, telecom penetration etc. For example, in many developed countries (for example, USA, Canada, UK), USO not only includes access to a PSTN, but also to directory services, selective outgoing call barring to premium service, emergency services and installation of payphones. In these countries, increasingly, USO also includes Internet access to public schools and libraries at discounted prices. In developing countries, the focus is on availability of at least a common telephone in rural areas. The emergence of Internet has led developing countries to incorporate access to it as a part of their USO. For example, in India, the National Telecom Policy 1999 laid down that by 2002 all villages would be provided with a telephone and all exchanges with reliable media. By 2010, teledensity in rural areas would be 4 and Internet access at all district head quarters would be provided as a part of USO. Based on the Telecom Development Report, 1998, the TRAI identifies the following criteria for USOs:

Availability: provision of telephone services whenever and wherever required even in remote and rural areas.

Accessibility: Non-discriminatory tariff in the service area regardless of the geographic location. Non-discrimination in terms of service quality, price.

Affordability: Telephone service to be provided at prices so that it is affordable to most users.

In developed countries the USO focus is on providing customer choice through competition. Moreover, since USO contributions as a percentage of the network revenues may not be a significant factor, USO policies may be designed differently. For example, Oftel estimates that "the real commercial cost of providing universal service at current levels is in the range of 0-0.5% of current network revenues. This relatively small overall figure makes it essential to devise a method for calculating and sharing any net costs of universal service which does not consume a disproportionate amount of resources (http://www.oftel.gov.uk/consumer/univ_1.htm#Foreword). In developing countries, this proportion could be much higher.

## Technology

While telecom costs have been falling dramatically, provision of local loop continues to be the most expensive part of the service provision chain. Wireless technology is fast emerging as commercially viable alternative for fixed line networks. This raises the issue of specification of technology in the fulfillment of USOs. The bundling of services such as cable/satellite or Internet may reduce the cost of telecom service provision. The regulatory challenge is to identify the USO provider and extent of USO support sought from service providers using different technologies.

## Reforms Process

Privatization in service provision also involves identifying the USO provider. In UK, British Telecom, the incumbent operator was identified as the USO provider. The privatization process needs to be viewed in the light of tariff re balancing in local and long distance that has been initiated in several countries as a part of the reform process. Dramatic reductions in the cost of long distance had allowed service providers to subsidize below cost local calls. But with the unbundling of services (such as separation of local, long distance and international) and competition in each segment, it has become imperative to increase local rates. In the process of re balancing, regulatory agencies need to ensure that services do not become unaffordable for certain segments. This may require targeted subsidies and thus has implications for funding and disbursal of USO.

*Frameworks for USO Funding*

Different mechanisms for USO funding are as follows:

1) An access charge mechanism: paid as interconnection charges by long-distance operators seeking access to the local loop. In this case all long distance operators pay an additional amount over and above the call terminating charges. In the US, this amount is referred to as the Access Deficit Charge and is calculated on the basis of the usage of the network, usually in terms of cents per minute. These amounts are separately entered on customer bills. The TRAI is also considering such a mechanism, especially in view of the opening up of the Long Distance services. Implementation of this mechanism requires a break up of revenues into local and long distance revenues.

2) Competitive bidding Regulators may choose specific auction mechanisms to give the service areas to the lowest/best qualified bidder who provides the stipulated services. Subsequently, on completion of the specified targets, the service provider is compensated for the amount that was bid (Nett, 98). Such mechanisms have been implemented in several countries including Argentina, Chile, Australia, and New Zealand.

3) License conditions, access to spectrum or other resources for private or government service providers could be bundled with build out requirements to ensure that USO are implemented. For example, In India 10% rural coverage in uncovered villages was imposed on basic fixed line service providers. In Philippines, licenses for profitable and unprofitable areas were bundled together. In addition, all licensees had obligations to provide a fixed percentage of their roll out in rural areas.

   Tradable USOs (Peha, 1999) attempt to take into account the inflexibility generally associated with implementation in terms of choice of technology and time periods. Recognizing that there are several impediments in fulfilling USO, this scheme suggests that USOs be made tradable. There would be two components to an obligation: the tasks to be done (milestones) and the time period within which these need to be completed (commitment).

   The model suggests that milestones and commitments be made tradable either by themselves or jointly. For example a long term commitment may be traded for a short term one. This would give flexibility in implementing the USO strategy of the operator depending on demand and technology choices available. The regulator could set a specified number of milestones each year, with commitments across all operators.

By having more milestones than commitments, least initial cost milestones will be met early. On the Commitment date, the firm can show it has met the milestone or it can trade its commitment with a longer term one, with some other party who has a more aggressive implementation plan. For this scheme to work, initially the regulator has to make all information regarding the milestones and commitments public, ensure that parties bidding for USOs are not just trading and ensure that in case of a failure of a party to implement a milestone, the penalty would be transferred to the previous owner. A similar scheme is operational in the US for pollution control where "permits" allow a company to emit a given amount of pollutants in the air. The number of permits are controlled and are tradable. In the context of USO, this is an innovative mechanism, giving flexibility to companies to take the dynamic ground realities into account. However, as of the time of this writing, there were no real life implementations.

**Rural Telecom Services in India**

As in many developing countries, India had poor telecom coverage. The Department of Telecommunications (DoT) was the monopoly service provider until 1995, when private services were licensed. The TRAI was established in 1997. Indian teledensity as of April 2001 was nearly 3.1, with rural teledensities being nearly one tenth of this. During the Eighth Five Year Plan (1992– 1997), the government decided that its objective of development would be accessibility rather than provisioning of individual phones. It, therefore, attempted to provide at least one pay phone per 100 households in urban areas and one pay phone per village. As a result, a large number of telecom booths, that provided local, long distance and international connectivity and often fax services sprang up across the urban areas. However, in the villages, the decision veered around to providing phones to the village panchayats[3] through a scheme called the Village Panchayat Telephone (VPTs) (Jain, 2000). Such phone services were not commercially oriented and often women and socially backward classes had little or no access to such facilities. This scheme was a failure since more than 50% of the installed phones did not work, large numbers did not have long distance calling and since there was no commercial incentives for villagers to ensure its working, several VPTs were disconnected due to non-payment (Jain and Sastry, 1997). While the availability of VPTs was poor, there was significant growth in rural services provided to individuals (table 1).

The private players in basic services were required to provide 10 per cent of their coverage in rural areas as a part of their license conditions. However, these requirements were not satisfied (TRAI consultation paper) and the DoT/TRAI had been able to do little about it. DoT's own record in rural service provision had also been dismal. Its rural coverage as a percentage of total coverage was 1.75 per cent. DoT has so far been claiming that its high cost of long distance was justified as it used the revenues from this service to subsidize rural services.

However, in 1998, its investment in rural services @Rs 75,000 per line are Rs 3210 million while its revenues in 1997-98 were Rs 163,650 million. Effectively, only 1.9% of it was invested in rural services and therefore, its justification for high long distance charge was weak.

---

[3] decision making bodies at the village level

**Table 1: Growth of Rural DELs and VPTs**

| Year | Rural DELs (millions) | VPTs |
|------|----------------------|---------|
| 1995 | 1.367 | 185,136 |
| 1996 | 1.761 | 211,113 |
| 1997 | 2.159 | 267,832 |
| 1998 | 4.2.55 | 310,687 |
|      |       |         |

Source: DoT's Annual Reports

A study that assessed the socio-economic benefits of rural telecom services (Jain and Sastry, 2000) showed that beyond voice services, there was significant demand for information services for which the villagers were willing to pay. Therefore, voice services bundled with other facilities like fax could become viable in sparse rural areas. The study also questioned the paradigm of rural obligations being loaded only on basic service providers[4].

DoT continues to identify basic services as necessarily fixed service. Cellular technology could also increasingly provide rapid coverage as costs of `cellular' fall more rapidly than that of fixed lines. Certain rural areas are more effectively covered with wireless (Jhunjhunwala, 2000). Rural obligations on the fixed service provider (FSP) become redundant if cellular service can cover rural areas. Convergence in service provision is increasingly making such issues more critical even in developed countries such as the USA (Garcia-Murillo and MacInnes).

**TRAI's Consultation Paper on USO**

Concerned about the poor coverage in villages and non fulfillment of obligations by private operators, the TRAI came out with a Consultation Paper on Universal Service Obligations[5] in July 2000 that considered a number of alternatives for assessing USO costs. While it recognized that detailed calculations would be more accurate, the cost of data acquisition for such an exercise would be very high.
For example, it considered calculating USO's at the local exchange level called the Short Distance Charging Area (SDCA)[6] since this could take into account local cost and revenue variations. But in view of DoT's inability to provide data at that level, suggested that in the future an approach based on categorization of SDCA could be adopted but for its paper it adopted average figures for VPT provision based on the DoT data.

---

[4] The government had mandated all private basic service operators to provide 10 per cent of all new lines in rural areas. A weightage of 15 per cent for service provision in rural areas was given at the time of bid selection. A penalty on a per day basis for each telephone not installed sought to prevent companies from delaying meeting rural targets

[5] http://www.trai.gov.in

[6] The accompanying box provides an outline of the network architecture.

This did not take into account the future costs of service provision, as the harder to reach villages had yet to be connected. The data that was provided by DoT covered only 5 of the nearly 21 states.

> Network Architecture
>
> For telecom administration the country has been organised into circles, each approximately corresponding to the state boundaries. The circles are further divided into Long Distance Charging Areas (LDCAs) or Secondary Switching Areas (SSAs). Each LDCA has an important town designated as the Long Distance Charging Centre (LDCC) for the purpose of charging long distance calls. For example, at the time of the consultation paper, a call travelling over 50 km was charged as a long distance call. The LDCAs are sub-divided into SDCAs, which approximately correspond to sub districts. For charging purposes, an important town in the SDCA is defined as the Short Distance Charging Centre (SDCC). There are 321 SSAs or LDCAs and around 2,550 SDCAs in the country. The charges are based on inter-LDCC radial distances for inter-LDCA calls and inter-SDCC for intra-LDCA calls.

The calculation of the net universal service cost (NUSC) i.e., USO revenues net of costs included varying estimates of the capital costs, capital recovery costs, operating expenses as a percentage of capital costs based on the different technologies available for VPTs (ie wirelines and analog Multi Access Radio Receivers (MARR)) and revenue estimates from existing and future VPTs. Recognizing that since the cost estimates of VPTs varied over a wide range, the TRAI paper used three values: Rs 50,000, Rs 75,000 and Rs 1,00,000 for building alternative scenarios. The revenue estimates took into account the distribution of phones with and without long distance facility. The sample covered nearly 10% of the existing VPTs. The estimates did not take into account that due to the poor technology characteristics of MARR, and high maintenance costs, these were often down and hence generated less revenue. But, in the future, these were unlikely to be deployed.

**Issues with the TRAI's Costing Approach**

*Sampling Issues*

The USO estimation was based on a study by consultants in 1998 and TRAI's own data collection exercise. The data on which the estimate was based is shown in table 2.

**Table 2: Sample Data**

| Circle | Sample VPTs – Local call facility | Sample VPTs – Long distance | Total sample |
|---|---|---|---|
| ICICI' sample covering 10 SSAs | 87 | 6 | 93 |
| Rajasthan | 18110 | 0 | 18110 |
| Bihar | 44 | 14 | 58 |
| Haryana | 1247 | 0 | 1247 |
| Tamilnadu | 27 | 0 | 27 |
| Karnataka | 18876 | 0 | 18876 |
| Weighted Average | 38391 | 20 | 38411 |

Source: http://www.trai.gov.in

It is clear that data from two states Rajasthan and Karnataka predominate. The consultants have pointed out the variations in cost reported by various exchanges. The explanation was that "the variation was mainly on account of improper allocation of "other" costs and due to different geographical terrain". Therefore, it was decided to work with "average capital expenditure at the national level" for USO estimates. This was because DoT did not have an accounting system that segregated costs based on rural and urban. This approach highlights the severe data constraints under which newly formed regulatory agencies operate.

*Validity of Assumptions*

Since cost variations of VPT provisions are high, it would be unreasonable to assume that average costs would be applicable. Such an approach would not create the right market incentives for the USO operators as it would possibly limit service provision in those areas that are high cost.

Moreover, given that the VPT scheme had been a failure, it is not clear why the TRAI took the revenue estimates of VPTs into account for calculating NUSC.
The new connections in rural areas are going to be commercial ventures in that the DoT has decided to install these phones in a commercial enterprise. Therefore, TRAI should have considered the existing VPT costs as sunk costs.

*Definition of Covered Villages*

Several studies have shown that at any point in time more than 50% of phones do not work and have not been working for sometime in the past. Several of the "covered" villages do not have long distance calling facilities. In this context, what should be the definition of "covered villages" which was taken into account to estimate NUSC needed to be spelt out in specific details.

**Deriving the Framework for USO Cost Estimation**

To arrive at a framework for USO cost estimation, we carried out a field study that covered ten exchanges in various villages of a district in Gujarat, India. The details of the exchanges are provided in table 3. To ensure representative elements of costs were assessed properly, the sample covered both large and small exchanges as well as variations in subscriber densities as shown by the variations in equipment capacity and subscriber densities in table 3.

*Size Variations*

These exchanges represented the full extent of size variations to be found among rural semi-urban exchange areas. Each exchange served a host of surrounding villages; the number varying from 4 to 27 villages and when we take the number of VPTs parented to this exchange, the number is larger.

*Subscriber Density Variation*

The variation of subscriber density is also quite high, ranging from 44 subscribers per sq. km in near urban conditions to less than 3 in small remote exchanges. The teledensity of the served areas are also seen to vary from less than 1 in the smallest exchange to nearly 5 for the second largest exchange in the sample.

**Table 3: Calculation of Local Loop Costs**

| Exchange Characteristics | | Characteristics of Exchange Area | | | Characterisctics of Villages served by the Exchange | | Parameters defining Local Loop Costs | | | |
|---|---|---|---|---|---|---|---|---|---|---|
| Name | Exchange Type | Equip. Capcity | Tele-density | Served Area | Served Population | No. of Villages Served by DELs | Max Distance (DELs) | Subscr. Density | CKM per Subscriber. | Cost per Line (Outdoor Plant) | % age Installation Cost |
| | | | | (Km Sq.) | | | (Km) | | | Rs. (000s) | |
| Nana Pondha | CDOT 128 | 88 | 0.7 | 31.7 | 13378 | 4 | 14 | 7.0 | 2.8 | 13.27 | 20.8 | 44% |
| Degam | CDOT 256X2 | 336 | 1.1 | 86.1 | 31763 | 7 | 265 | 14.0 | 3.9 | 13.11 | 20.3 | 42% |
| Rumla | CDOT 256X2 | 336 | 0.9 | 75.9 | 35555 | 9 | | 7.5 | 4.4 | 9.39 | 15.6 | 42% |
| Mota Wagchhipa | SBM 512 | 360 | 1.0 | 68.4 | 36961 | 16 | 183 | 13.0 | 5.3 | 11.16 | 13.9 | 31% |
| Kheralu | CDOT 1 K + 512 | 1280 | 3.4 | 81.4 | 37987 | 13 | 146 | 7.5 | 15.7 | 6.41 | 9.3 | 34% |
| Bansda | CDOT 1.4 K | 1400 | 2.4 | 127.9 | 58456 | 23 | 159 | 13.0 | 10.9 | 9.14 | 12.8 | 36% |
| Dharampur | CDOT 1.4 + 256 | 1552 | 1.5 | 223.1 | 102861 | 27 | | 10.0 | 7.0 | 9.80 | 11.9 | 28% |
| Killa Pardi | CDOT 1.4 + 512 | 1860 | 4.1 | 42.6 | 45203 | 4 | | 5.0 | 43.6 | 5.96 | 5.9 | 19% |
| Sarigam | CDOT 1.4+1 K | 2432 | 4.9 | 95.6 | 49526 | 13 | 793 | 12.0 | 25.4 | 9.34 | 9.6 | 24% |
| Chikhli | CDOT 1.4+1.5 K | 2992 | 3.7 | 120.6 | 81290 | 16 | 809 | 9.0 | 24.8 | 6.11 | 6.5 | 21% |
| AVERAGE | | 1264 | 2.6 | 95.32 | 49298 | 13 | 338 | 9.8 | 13.3 | 8.00 | 9.6 | 29% |

**Assessment of Costs**

The components of the cost per line are outdoor plant, switching and transmission; we have focussed on the first of these components namely the per line cost of the outdoor plant or the local loop which is the cost of the equipment / network from the exchange to the subscriber premise.

This is the single largest cost component subject to wide variations especially in rural networks and we have attempted to correlate this cost to the size of the exchange and subscriber density - the two most obvious, though inter-related, dimensions. All cost calculations pertain to the land line technology, as the existing MARR systems hardly worked and are not to be adopted for future spread.

In the village exchanges, we made detailed cost estimates of the outdoor plant layout, both primary and distribution segment in terms of each cable section, gauge and size (number of pairs). Material costs and installation costs were separately calculated for each cable section and then summed up. The item-wise material costs per unit used were obtained from the DoT for last financial year. The cable cost and cost of other material like jointing kit, C T boxes, cabinets, pillars etc are constant over the sample; but since the DoT is the monopsonist bulk buyer of these items, it is possible that these prices may not reflect the market price and to that extent there might be some underestimation in our calculations.

The estimated cost per line of the outdoor plant is shown for each exchange area. It is observed that there is a lot of variation in installation costs since (1) digging and trenching costs depend on the hardness of the soil and (2) the cost of labour and both these vary from place to place. On an average installation costs are 30% of total costs.

The per line cost is as small as nearly Rs. 6,000 for Killa Pardi - the largest exchange with nearly 3 K line, and as high as 21,000 for the small CDoT 128 port exchange in Nana Ponda. The average local loop cost is around Rs. 9600. Some of the villages served by the exchanges are situated as far as 14 km[7]; consequently the local loop costs are far greater than compact urban service areas. The average distance of the subscriber is reflected in the CKM (conductor kilometer) per line that is simply twice the average distance; the CKM per line is seen to vary from 13.2 km for a small exchange area to a 6 km for a large exchange

**Analysis of Data**

*Loop Cost variations with Exchange Size*

Analysis of the loop cost variation with exchange size shows that two distinct groups exist. In one group we have the exchanges which are 512 port (CDOT) size and less; for this group the local loop cost averages around Rs. 20,000. The other group is of size 1 K lines and above; these are the SDCA level exchanges and the average loop cost is Rs.10, 000 (table 3, Figure 1).

---

[7] This might sound improbable but typically what happens is that part of the loop will be buried cable (smaller gauge 0.5 mm) and the last few kilometers will be carried on overhead GI wires which have a higher gauge and minimize the loss to some extent.

Figure 1: Exchange Size and Local Loop

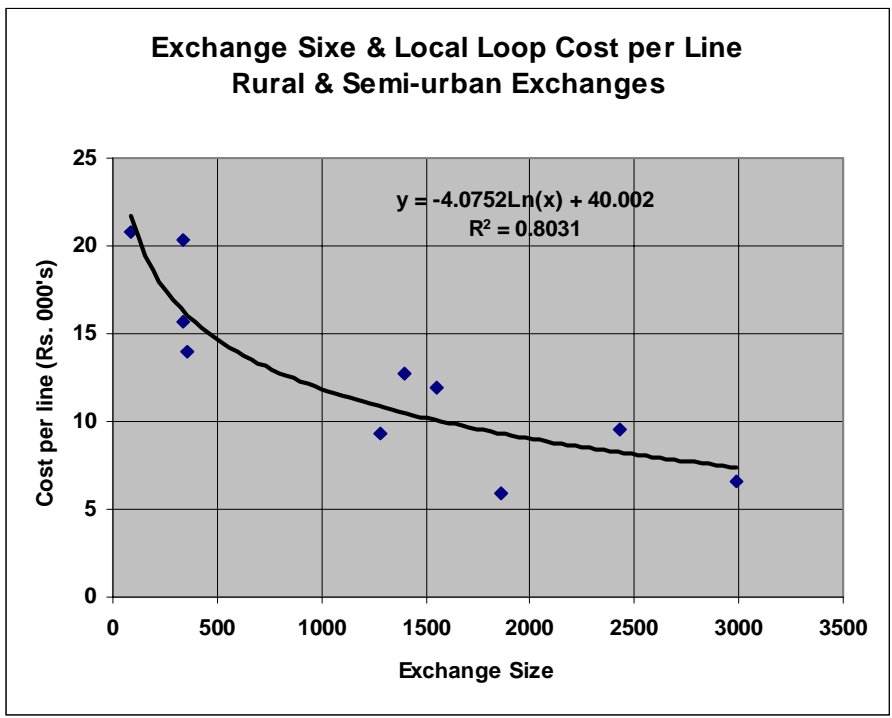

The variation in terms of the size of the exchange is not always accurate. It is possible to have a small rural exchange serving a compact small area , may be a few villages around. This will be a case of small size and low costs. Again we could have a large exchange serving a thinly spread out areas, this will lead to the case of large size and large costs per line. Both these cases will go against the average notion of per line cost varying inversely as size. These cases are not anomalies but in part a consequence of the fact that most rural (growing) exchanges go through different phases in their life cycle. The initial stage of a rural exchange is essentially compact service area in the immediate vicinity. Then with time and demand growing in the adjoining areas, the outdoor plant spreads to more and more distant villages. This, however, cannot go on indefinitely because the physical factors of attenuation consequent on large resistance of the loop. Depending on the subscriber density of the region there is a definite optimal size beyond which a rural exchange cannot grow

*Loop Cost Variations with Subscriber Densities*

We examined the subscriber density associated with an exchange. (Policy matters typically involve use of teledensities, but this is not a very useful category for our purposes since the same level of average subscriber density may correspond to various levels of teledensity depending on the population density of the region. (Subscriber Density = Teledensity * Population Density)). We also present the variation of subscriber density with exchange size in figure 2.

Figure 2: Subscriber Density and the Cost of the Local Loop

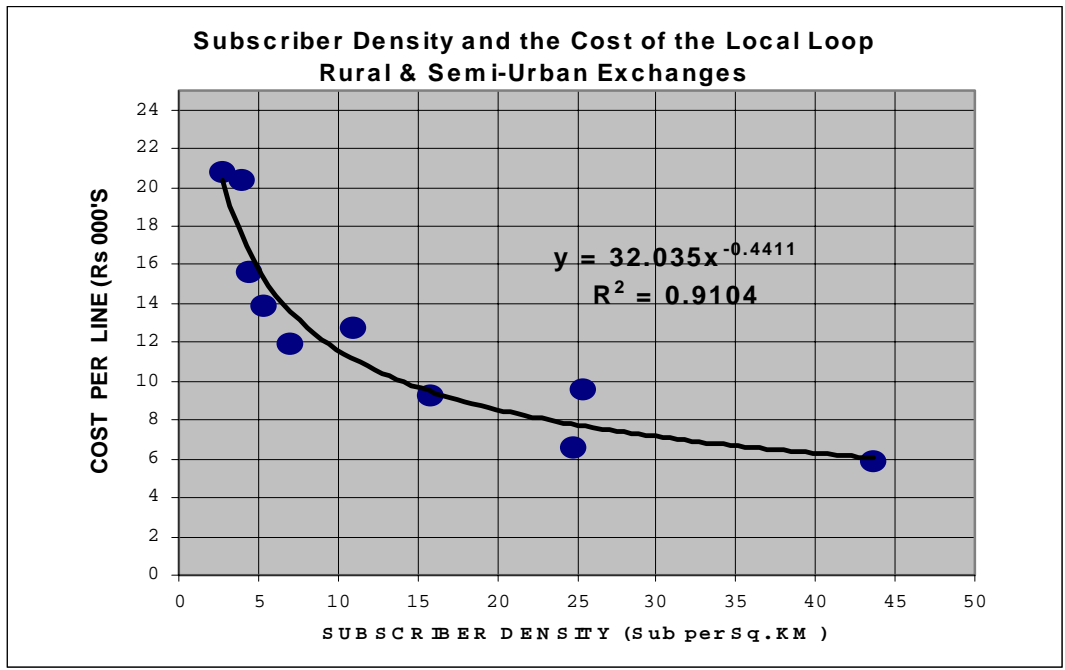

The small exchange with a compact serving area will show high density and low cost; similarly the spread-out big exchange will show low density high cost. It is thus clear that it is the subscriber density that significantly affects the cost of the local loop.

We have, therefore, investigated this relationship in our sample. The double-log function fits the scatter very closely as seen in the following figure.

**Ln ( Cost per line )  =  3.467  -  0.4411 Ln ( Subscriber density )...........R-square = 91%**

( 28.94)        ( -9.02)

Figures in parenthesis represent the t-statistics of the estimated regression coefficients.

The plotted points can be grouped in three clusters (a) those which have subscriber density below 5 and costs averaging above Rs.20,000 per line, (b) those with subscriber density between 5 and 10 have costs averaging around Rs.14,000 and (c) those with density above 10 in the sample average costs are in the range of Rs.8,000 per line.

The variations in costs are brought about by the opportunities of common cost sharing. The most important common costs are that of trenching, laying, back filling and material protection. Thus for small rural exchanges installation costs are as high as 44% of the total cost. In comparison the higher density areas show a 20% installation cost share. The advantages of common costs increase as we move on to bigger exchange areas with higher subscriber densities resulting in falling installation costs as a proportion of total costs. Beyond a point - in our case beyond a density of say 50 per sq. km - the gains from increased density on account of common costs fall off progressively and the cost per line tends to flatten out.

Further, we present the variation of subscriber density with exchange size for one SSA with nearly 90 exchange areas (figure 3). The size varies from 30K urban exchanges at the SSA headquarter to the smallest rural exchange. The relationship is found to fit the linear case very closely.

**Subscriber Density  = 0.0179  * (Exchange Size)  + 0.0169..........R-Square = 98.8%**
                        (87.6)                      (0.2)

It is found that for every hundred lines in an exchange area the subscriber density increases by approximately 1.8 per sq. km. Our sample clearly is drawn from the less than 50 subscriber ® density region which accounts for most of the exchange areas in any typical SSA in the Indian telecom ??

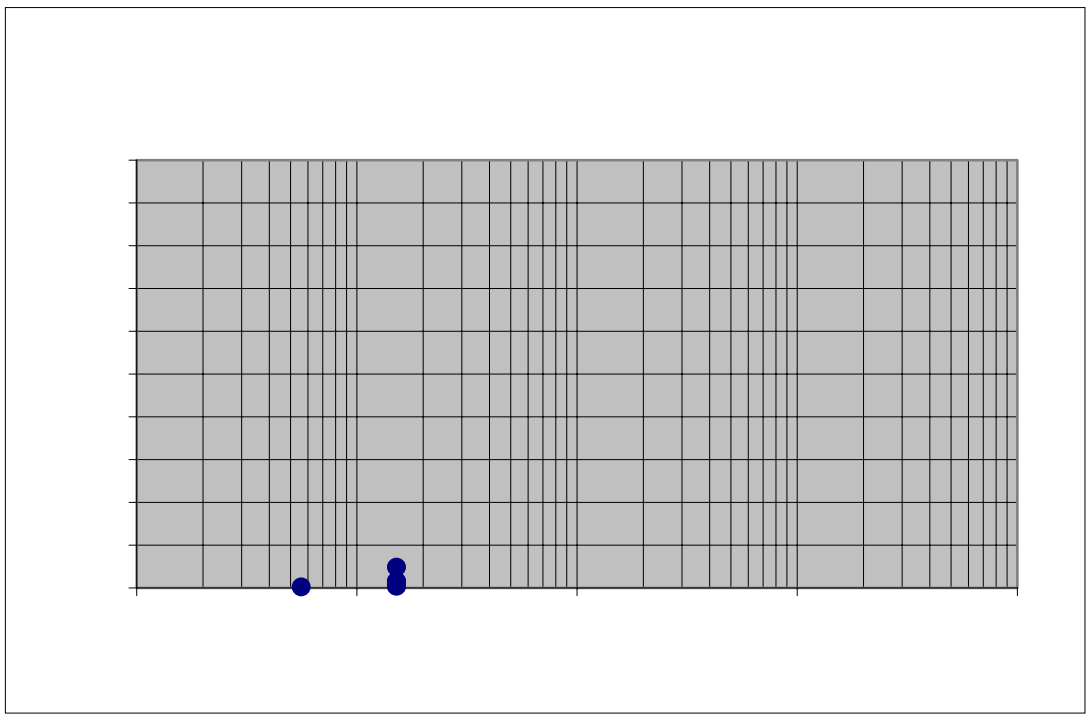

Figure 3: Subscriber Density and the Size of the
e

*Cost Variations based on Capacity Utilization*

In part, the actual local loop costs would also vary with capacity utilization. However the problem was not too severe for the Gujarat Circle where rural and overall capacity utilization is very high; further, the recent reductions in the deposit money for new connections has resulted in a huge surfacing of latent demand, especially in the rural

Since our sample exchanges had varying capacity utilization levels and since outdoor plant is planned to capacity we have taken the equipped capacity as the denominator in all our calculations. In those areas where the capacity utilization is not high, this could result in an underestimation of smaller exchanges area costs because typically newly commissioned small rural exchanges tend to remain under utilized for a long time. The capacity utilization of an exchange depends on a lot of factors such as (1) demand and (2) the minimum quantum of expansion that is technically

*Rationale for Choosing SDCA as the Unit of s*

While calculating USO costs, there is a tradeoff between costs of detailed assessments and variations in costs across the chosen units with those specified by regulatory agencies. We suggest that thappropriate unit of disaggregation for USO purposes should be the SDCA. Thus, for example, population densities and labor and other costs of installation, which are a high proportion of local loop costs would not vary significantly within SDCAs but could vary between SSAs. Beyond the SDCA level, it would be difficult to control for variables that would significantly effect the cost parameters.

Typically, an SDCA will have a mixture of exchange sizes, data on which is readily available. From this distribution of exchange size one could derive the average subscriber density with the help of the density - size relationship presented above, to arrive at the SDCA level subscriber density and relate this to the average local loop cost. Further, the SSA could be categorised based on existing telecom network, population densities, terrain features and other demographic information. This involves carrying out the analysis for a typical SDCA level within an SSA, for all SSAs. Since the number of SSAs is around 300, the costs of computation are not high. Further, even among SSA's if some groupings can be made on the basis of similarity in terms of geographic features and population characteristics, then lesser work needs to be done. Existing Geographic Information Systems packages could be used effectively to do the groupings.

Our approach also takes into account the life cycle effects of subscriber growth in an exchange area. By correlating the growing subscriber density with the local loop costs, estimates for USO costs may be revised periodically. In the absence of a model like ours, the estimates would have to be done afresh whenever USO costs needed to be reviewed.

*Need for dis aggregation*

There is a clear need to dis aggregate the type of exchanges for cost and revenue estimation. There is for instance, a tendency to speak of "rural" exchanges or "urban" exchanges as if each were an amorphous lot – all exhibiting the same characteristics – and hence justified to be clubbed together. While this facilitates calculations, it leads to too much aggregation and consequent implications of ill-directed subsidies. It is obvious that for a network size such as ours, it should make no sense to direct subsidies based on one or two cost per line or one or two revenue per line figure.

*Basis for Classification of Exchanges*

In India, there is some confusion on what is rural and urban because the Census Bureau and the DoT have chosen to define their own set of criterion. As per DOT criteria, 1 K is as the cut-off exchange size for an exchange to be classified as rural. Census Bureau has its definitions based on workforce and other characteristics. These two definitions lead to a cluster / set of small towns and big villages which have dual status. We have hence chosen a sample of exchanges of 3K lines and less; and these should cover the rural-semi-urban spectrum. Ideally, for the purpose of USO and for telecom policy purposes in general, the definition ought to be in accordance with easily observable characteristics of the settlements and its inhabitants; characteristics that relate to cost of provision and its returns.

The obvious candidates are population or subscriber density and income or some related variable that are tracked at the appropriate level. Only then can regulators achieve greater transparency and clarity in directing subsidies to target groups.

*Conclusions*

This paper has attempted to provide a model of cost estimation for the local loop for estimating USOs. This model did not take into account the exchange and transmission costs associated with the telecom service. Since this could be done at a more broad level, and is not likely to vary as much with distance, terrain characteristics, and socio-economic factors, it may not lead to as much contention.

Since regulators in most developing countries grapple with non-availability of data, this simple approach could help them to minimize the constraints they face.

While TRAI's approach to USO estimation was driven by an appropriate understanding of the factors that influence USO, its adoption of values provided by DoT, may lead to concerns regarding the credibility of the exercise. Its estimates of revenues based on existing patterns that are derived from the VPT scheme that is considered a failure was flawed.

In such a fast changing technological scenario, it would be difficult for any regulatory agency to "get the technology right". TRAI's recommendation should therefore be technologically neutral. As regards the cost of technology, a disaggregation at SDCA level by service area characteristics embedding both socio-economic conditions and technology characteristics should be done. Payments should be based on bidding and close monitoring of milestones.